\documentstyle[11pt,aaspp4,twoside,psfig]{article}
\index{pulsars:individual}
\index{supernova remnants: individual}
\index{star: individual}
\index{stars: neutron}
\index{X-rays: general}

\def\lum{$\rm D^{2}_{10 \rm\  kpc}$ ergs s$^{-1}$}

\def\edcomment#1{\iffalse\marginpar{\raggedright\sl#1\/}\else\relax\fi}
\reversemarginpar

\begin{document}
\title{Detection of X-ray Emission from SNR G16.7+0.1}
\author{Marcel A. Ag\"ueros}
\affil{Astronomy Department, University of Washington, Box 351580, Seattle, 
WA 98195, USA}
\author{David J. Helfand}
\affil{Columbia Astrophysics Laboratory, Columbia University, 550 West 
120$^{th}$ Street, NY, NY 10027}
\author{Eric V. Gotthelf}
\affil{Columbia Astrophysics Laboratory, Columbia University, 550 West 
120$^{th}$ Street, NY, NY 10027}

\begin{abstract} We have observed the Galactic supernova remnant G16.7+0.1
for 13 ks using the EPIC cameras aboard the {\sl XMM-Newton} X-ray Observatory,
producing the first detection of the SNR outside of the radio band. G16.7+0.1 
is one of the faintest radio synchrotron nebulae yet detected, although the 
core-to-shell flux ratio at 6 cm is typical of composite SNRs. 
The distance to the object is unknown. Our image is seriously 
contaminated by single-reflection arcs from the X-ray binary 
GX17+2, which lies just outside the field of view. Nonetheless, the 
remnant's synchrotron 
core is clearly detected. We report on the spectrum and intensity of the core 
emission as well as on our search for emission from the thermal shell, 
and describe the constraints these observations place on the SNR's 
distance, age, and central pulsar properties.  \end{abstract}

\section{Introduction}

X-ray observations of SNRs provide crucial information on the poorly known
initial distribution of pulsar magnetic field strengths and spin
periods, as well as on the ages and dynamics of the supernovae which
created them. To date, only five Galactic Crab-like and eight composite
remnants have been detected at X-ray wavelengths; often, 
the essential spectral information is poor.

\begin{figure}
\centerline{\psfig{figure=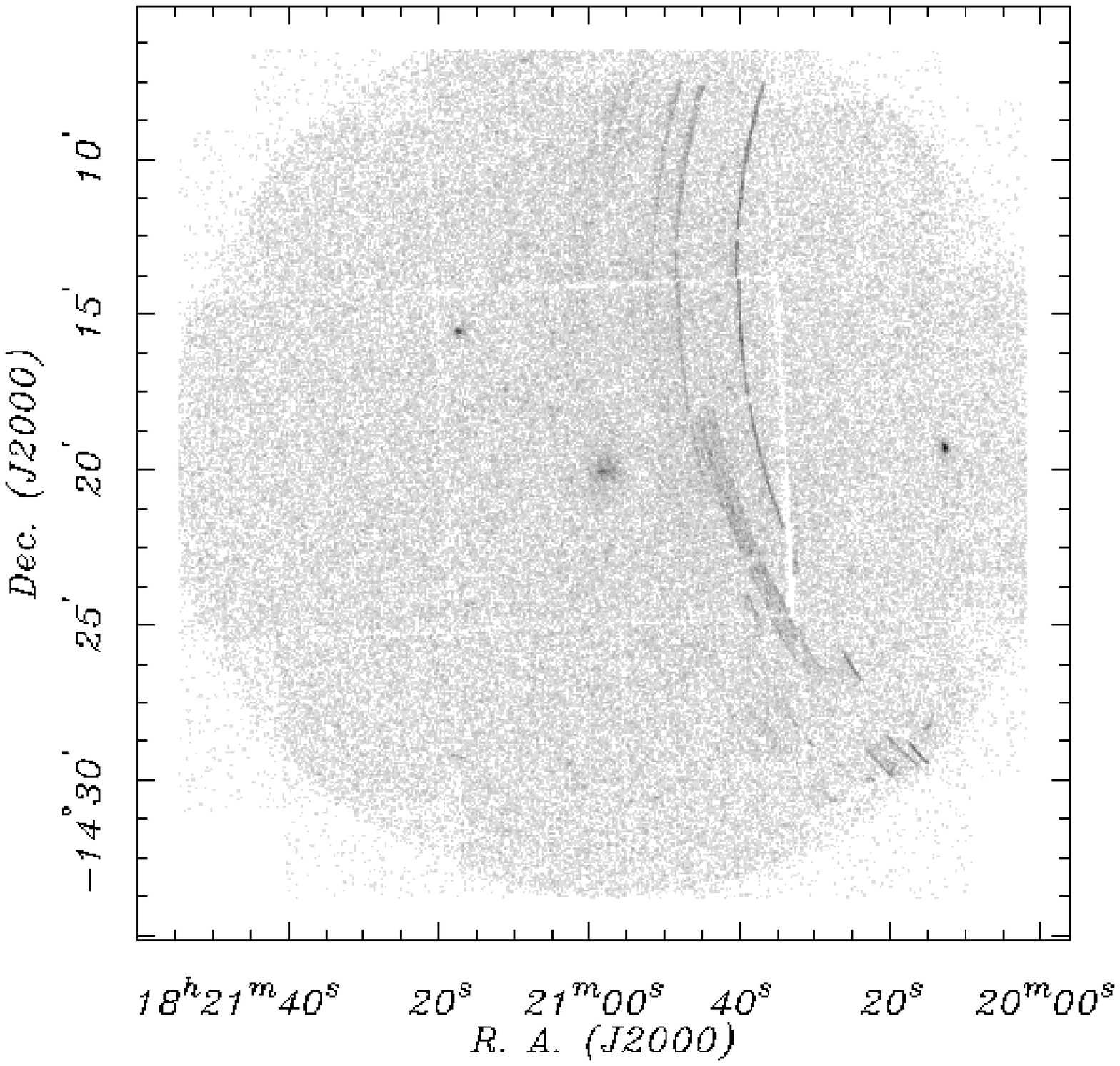,height=4.0in}}
\caption{The merged MOS1 plus MOS2 image of SNR G16.7+0.1. The image is displayed with logarithmic intensity.}
\label{picture}
\end{figure}

A classic composite remnant with the faintest radio core component detected 
to date (only 100 mJy; Helfand et al. 1989), G16.7+0.1 has
comparable radio luminosities in the core and shell components; 
the shell and core radii are $\sim 2\arcmin$ and $\sim 1\arcmin$, 
respectively. G16.7+0.1 was observed on $8-9$ March 2001 with the
{\sl XMM-Newton} Observatory. Data were obtained
from all three of the cameras which form the EPIC instrument 
(see Figure \ref{picture}). The PN and the two MOS cameras  
are CCD arrays sensitive to photons with energies between
0.1 and 15~keV; the CCD pixel sizes are 1.1 and 4.1$\arcsec$, respectively.

\section{Method}

The {\sl XMM} Standard Analysis System (SAS) was used 
to filter the data for good events within the camera's energy 
range; additionally, periods of high background activity
were identified and removed. This produced filtered data sets of
8.7~ks and 8.8~ks for the MOS cameras, and of 5.5~ks for the PN camera.

Because of its distance and position, the absorption column density ($N_H$) to
the remnant is high, and few source photons are detected at energies
$\leq 1$~keV. Spectral fitting was done
from 1 to $\sim 10$~keV, with the upper limit
being slightly different for each camera. The position of
the remnant on the PN camera was perilously close to the gap between
two chips. These data were therefore used exclusively to verify the 
consistency of results obtained from MOS data.

The bright arcs easily visible in the data from all three detectors
are due to single-scatter photons from the nearby bright X-ray binary GX17+2.
To characterize the contribution of the arcs to the source
background, where they are dim but present, we compared 
spectra of different arc regions
to detect possible variations. Finding none, we divided the central MOS
chip into three polygonal areas. This geometry mimics the elliptical shape
of the arcs; a given
region contains a full section of any arc present in that part of the chip.

\begin{figure}
\centerline{\psfig{figure=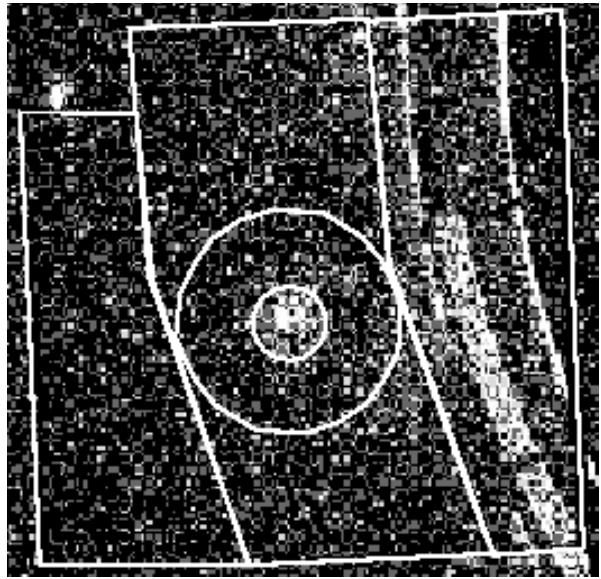,height=3.0in,clip=}}
\caption{The regions used in analyzing the filtered MOS1 data. The outer boundaries correspond to the central chip edges.}
\label{region}
\end{figure}

Figure \ref{region} is a section of the filtered MOS1 image. The area
to the right is used to characterize the arcs, the one
to the left to characterize the ``normal'' background. The central area
corresponds to the source background and is a weighed sum of arcs and
background. Two circles are centered on the source; the larger
(R = 135\arcsec) is the radio remnant, while the smaller (R = 45\arcsec)
zeroes in on the brightest emission
region, coincident with the radio synchrotron core. The circles' center,
RA $18\fh20\fm57\fs8$, DEC $-14\fdg20\farcm09\farcs6$, 
is the SNR's apparent geometrical center at 6 cm (Helfand et al. 1989). 
The different areas are not equal and scalings are applied 
in correcting or comparing them.

All spectra were grouped to contain a minimum of 25 counts bin$^{-1}$, and
locally generated response functions were employed.
The canned response files, available from the
{\sl XMM} website, were used to check for consistency. 
All errors are quoted for a 1-$\sigma$ confidence range.

\section{Analysis and Results}

\subsection{The X-ray core}

Spectra were extracted from the MOS images for a circular region of radius
45\arcsec, and fit simultaneously over the energy ranges 1 to 8.4 keV for MOS1
and 1 to 7.5 keV for MOS2 (the upper limits being set by the data quality).

\begin{figure}
\centerline{\psfig{figure=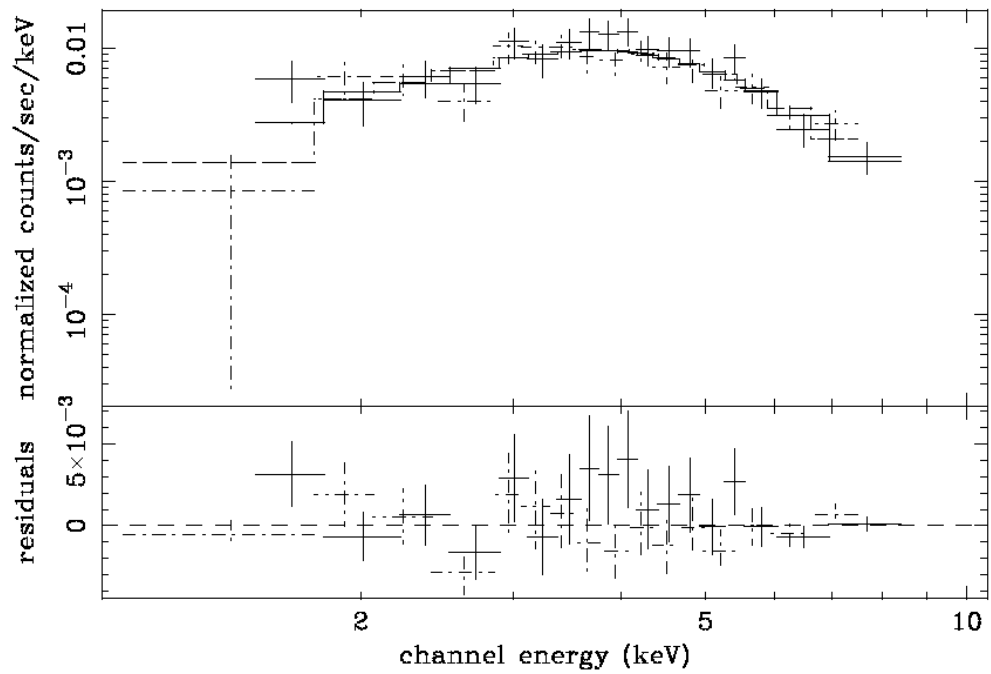,height=3.5in,angle=270}}
\caption{Fitting the MOS1 (solid points and line) and MOS2 (dot-dashed points and line) spectra simultaneously.}
\label{fit}
\end{figure}

A power-law modified for interstellar absorption is an
excellent description of the core's spectrum. XSPEC returns
a photon index $\Gamma=1.17 \pm 0.29$ and 
$N_H=4.74 \pm 0.98 \times 10^{22}\ {\rm cm^{-2}}$,
with a reduced $\chi^{2}_\nu=0.82$
for a fit using the backgrounds models 
and locally generated RMFs and ARFs (see Figure \ref{fit}; the results
obtained with the canned response files are essentially identical, with
$\chi^{2}_\nu=0.84$). Fixing $\Gamma$ and
$N_H$ to these values and fitting the extracted PN spectrum (for a
smaller area: R = 28\arcsec) returns $\chi^{2}_\nu=0.79$ over
1 to 10 keV.

For a soft band of 1 to 3 keV and a hard band of 3 to 8 keV, the hardness 
ratio ((N$_{hard}-$N$_{soft}$)/(N$_{hard}+$N$_{soft}$); N is the number 
of counts) of the core for the two MOS cameras averages 0.56.

In the direction of G16.7+0.1, the total Galactic HI $N_H =
1.56 \times 10^{22}\ {\rm cm^{-2}}$. Generally, X-ray derived 
$N_H$ values are $\sim 2$ to 3 times the radio $N_H$ for a given object.
The SNR is thus at least on the other side of the
Galactic Center; we take D $=10$ kpc.
PIMMS gives an unabsorbed flux for the core of 
$1.9 \times 10^{-12}$ ergs cm$^{-2}$ s$^{-1}$ over 0.5
to 10 keV, so $L_{Xc} = 2.3\times 10^{34}$ \lum.

G16.7+0.1's core has a radio luminosity $L_{Rc} = 0.3 \times 10^{34}$ ergs s$^{-1}$,
so $L_{Xc}/L_{Rc}=8$, similar to MSH15-52
($\geq10$) and CTB80 ($\sim3$) (Helfand \& Becker 1987), and 
intermediate between very young SNRs ($\sim100$ for Kes 75, PSR B0540-69, and 
the Crab), and older SNRs ($\sim1$ for Vela).

\subsection{The remnant shell}
The small number of counts ($\sim100$ for each MOS camera)
from the SNR shell (defined as coincident with the radio nebula's area,
R $=135\arcsec$) does not allow us to constrain the nature of the 
emission or comment on its hardness ratio relative to the core's. 

If D $=10$ kpc, R$_{shell} = 6.5$ pc and R$_{core} = 2.2$ pc. 
The distance-independent ratio R$_{core}$/R$_{shell} = 0.3$, 
consistent with ratios found from 6 cm observations of other
composite SNRs (Helfand \& Becker 1987). If G16.7+0.1 
is in the free-expansion phase with $v_{exp} \approx 3 \times 10^3$ 
km~s$^{-1}$, it is $\sim 2100$ years old.

\subsection{A central pulsar}

Seward \& Wang (1988) obtain pulsar characteristics
from the SNR $L_{X}$ over the 0.2 to 4 keV range.
In this range, $L_{X} = 1.1 \times 10^{34}$ 
ergs s$^{-1}$; the unseen pulsar's
$\rm \dot{E}=2.7 \times 10^{36}$ ergs s$^{-1}$. If the pulsar is indeed 2100
years old, P~$\sim 0.3$ s, $\rm \dot{P} \sim 2500 \times 10^{-15}$ s s$^{-1}$,
and $\rm B_0 \sim 30 \times 10^{12}$ G. 

\acknowledgements
D.J.H. acknowledges support from grant NAG5-9928. E.V.G. is supported by 
NASA LTSA grant NAG5-7935. We thank the {\sl XMM} helpdesks for their assistance.

\end{document}